\documentclass[aps,prf,showpacs,floats,superscriptaddress,shortbibliography]{revtex4-2}

\usepackage{wasysym}
\usepackage{array}
\usepackage{hyperref}
\usepackage{stmaryrd}
\usepackage{tikz}
\usepackage{float}
\usepackage{graphicx}
\usepackage[english]{babel}
\usepackage{graphics}
\usepackage{xcolor}
\newcommand{\TS}[1]{#1}

\date{\today}

\begin{document}

\title{Viscous to inertial coalescence of liquid lenses: a lattice Boltzmann investigation}
\author{Thomas Scheel}
\email{t.scheel@fz-juelich.de}
\affiliation{Helmholtz Institute Erlangen-N\"urnberg for Renewable Energy (IEK-11), Forschungszentrum J\"ulich, Cauerstr.\,1, D-91058 Erlangen, Germany}
\affiliation{Department of Physics, Friedrich-Alexander-Universit\"at Erlangen-N\"urnberg, Cauerstr.\,1, D-91058 Erlangen, Germany} 
\author{Qingguang Xie}
\email{q.xie@fz-juelich.de}
\affiliation{Helmholtz Institute Erlangen-N\"urnberg for Renewable Energy (IEK-11), Forschungszentrum J\"ulich, Cauerstr.\,1, D-91058 Erlangen, Germany}
\author{Marcello Sega}
\email{m.sega@ucl.ac.uk}
\affiliation{Department of Chemical Engineering, University College London, London
WC1E 7JE, United Kingdom}
\affiliation{Helmholtz Institute Erlangen-N\"urnberg for Renewable Energy (IEK-11), Forschungszentrum J\"ulich, Cauerstr.\,1, D-91058 Erlangen, Germany}
\author{Jens Harting}
\email{j.harting@fz-juelich.de}
\affiliation{Helmholtz Institute Erlangen-N\"urnberg for Renewable Energy (IEK-11), Forschungszentrum J\"ulich, Cauerstr.\,1, D-91058 Erlangen, Germany}
\affiliation{Department of Chemical and Biological Engineering and Department of Physics, Friedrich-Alexander-Universit\"at Erlangen-N\"urnberg, Cauerstr.\,1, D-91058 Erlangen, Germany} 
\begin{abstract}
	Liquid lens coalescence is an important mechanism involved in many industrial and scientific applications. It has been investigated both theoretically and experimentally, yet it is numerically very challenging to obtain consistent results over the wide ranges of surface tension and viscosity values that are  necessary to capture the asymptotic temporal behavior in the viscous and inertial limits. We report results of massively parallel simulations based on the color gradient lattice Boltzmann method, which overcome these limitations, and investigate the scaling laws of both regimes. For the two-dimensional case we find good agreement with the similarity solution of the thin-sheet equation, where in the viscous regime the connecting bridge grows linearly with time and in the inertial regime proportionally to $t^{2/3}$. In three dimensions, the viscous growth of the bridge also exhibits a linear time dependence, while in the inertial regime the growth of both the bridge height and the bridge width is proportional to $t^{1/2}$.
\end{abstract}

\maketitle
\section{Introduction}
From the formation of raindrops~\cite{Hu-Srivastava1995} to biomolecular condensates during liquid-liquid phase separation~\cite{Garaizar2022}, drop coalescence plays an important role in many natural phenomena, but finds also broad industrial applications. 
The latter include, for instance, sintering~\cite{Martinez-Herrera-Derby1995,Eggers-Lister-Stone1999},
filtration~\cite{Yarin-Reneker2006,Bansal-Planck2011}, and ink-jet printing of a variety of materials~\cite{WIJSHOFF201820,Hack-Karpitschka-Snoeijer2018,Sun2015} ranging from solar cells~\cite{Eggenhuisen2015,Eggers-Paetzold2020,MAISCH2021305,RH22a} to bioengineered tissues~\cite{Detsch-Boccaccini2016, Ruiz-Alonso-Pedraz2021} and cells~\cite{Kumar-Zhao2021}. 
Future improvements in these technologies rely strongly on the ability to advance the understanding of the wetting behavior of droplets on liquid substrates as well as an accurate knowledge of the interaction between the liquid phases and the dynamics of their coalescence~\cite{Lohse2022, Kuang-Song2014-2}. 
Previous research has primarily been focused on the coalescence of freely suspended droplets \cite{Eggers-Lister-Stone1999, Duchemin-Josserand2003, Aarts-Bonn2005, Paulsen-Nagel2011, Paulsen-Basaran2012, Sprittles-Shikhmurzaev2014, Zimmermann-Zeiner2020}
and droplets on solid substrates~\cite{Menchaca-Rocha-Zaleski2001, Ristenpart-Stone2006, Narhe-Pomeau2008, Lee-Yarin2012,Eddi-Snoeijer2013,Kaneelil-Stone2022}, while droplets on liquid substrates~\cite{Burton-Taborek2007, Hack-Snoeijer2020, Chen-Yin2022} have received less attention. 

The theoretical analysis of the coalescence of liquid lenses, i.e.~droplets attached to a fluid-fluid interface, has identified two distinct dynamic regimes, which depend on the relative importance of viscous and inertial forces~\cite{Aarts-Bonn2005}. Immediately after two droplets get in contact, inertial forces are still small compared to viscous ones, and the connecting meniscus height $h_0(t)$ is reported to grow proportionally to the elapsed time $t$~\cite{Burton-Taborek2007,Paulsen-Nagel2011,Eddi-Snoeijer2013,Hack-Snoeijer2020}. This linear dependence marks the so-called viscous regime. At longer times (or for larger surface tension to viscosity ratios), coalescence enters the inertial regime where viscous forces become negligible, and $h_0(t)$ is reported to grow like $h_0(t) \sim t^{2/3}$ for low contact angles $\theta \ll 90^{\circ}$~\cite{Eddi-Snoeijer2013,Klopp-Eremin2020,Hack-Snoeijer2020}.
The case of contact angles close to $90^{\circ}$ (as encountered in freely suspended droplets) turned out to be a particular one~\cite{Eddi-Snoeijer2013}, where a scaling $h_0(t)\sim t^{1/2}$ is often reported~\cite{Duchemin-Josserand2003,Aarts-Bonn2005,Burton-Taborek2007,Paulsen-Nagel2011,Eddi-Snoeijer2013,Xia-Zhang2019}.

\begin{figure}
\begin{center}
    \includegraphics[width=1 \columnwidth]{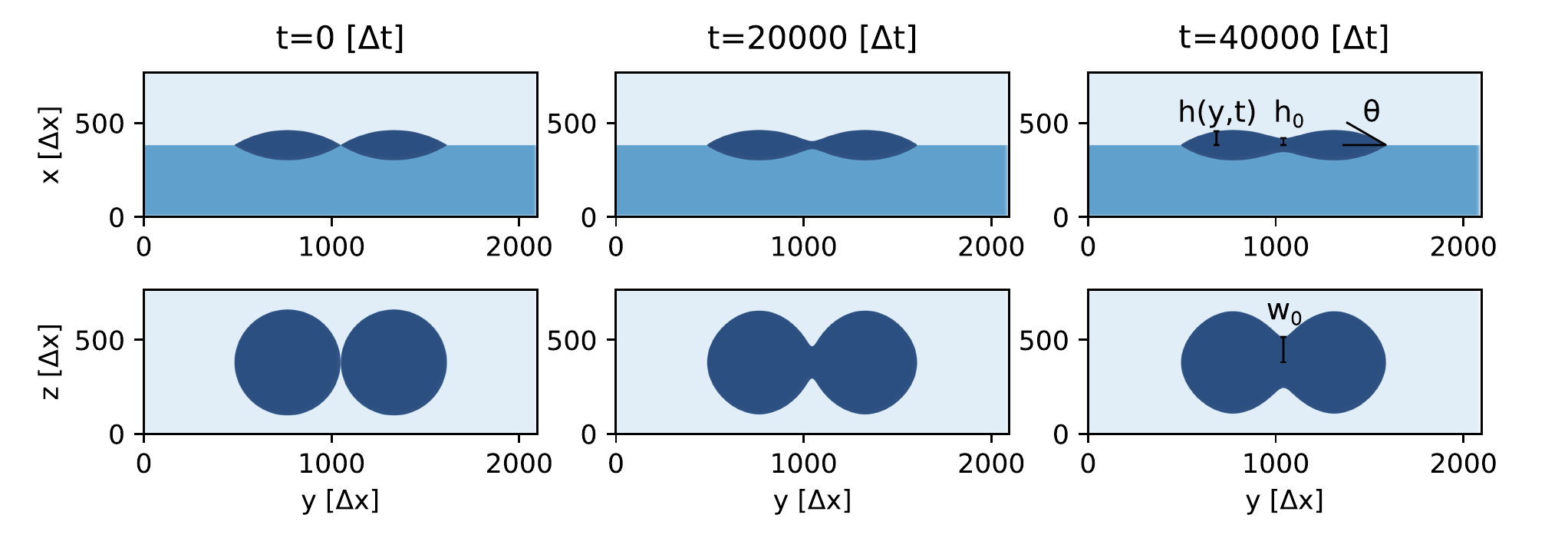} 
    \caption{Simulation snapshots of 3D liquid lens coalescence in the \TS{$y-x$} plane (top row: side-view) and in the \TS{$y-z$} plane (bottom row: top-view) at different simulation times $t$ with contact angle $\theta$, lens height $h(y,t)$, minimal bridge height $h_0$ and minimal bridge width $w_0$.
    Lengths and times are provided in lattice Boltzmann units $\Delta x$ and $\Delta t$. 
    }
    \label{fig:profiles3d}
\end{center}
\end{figure}
From an experimental point of view it is very challenging to resolve the coalescence process in sufficient detail at least for low viscosity liquids such as water. Here, it is practically impossible to observe the viscous regime, because inertial effects become dominant at the scale of $h_c \approx 15$~nm and for times larger than $t_c\approx10^{-10}$~s~\cite{Aarts-Bonn2005}.
On the other hand, analytical approaches rely on assumptions such as a reduced dimensionality (thin-sheet equation) or infinite bridge growth. The difficulties involved in the experimental measurements and the approximations used in the analytical treatments have prevented an unambiguous understanding of the scaling laws of three-dimensional liquid lenses with arbitrary wetting properties.

In order to describe the growth dynamics of top-down symmetric liquid lenses (see Fig.~\ref{fig:profiles3d}) it has to be taken into account that they involve two principal radii of curvature. As a result, the bridge is characterized not only by its height $h_0$, but also by its width $w_0$. Heuristically, one can imagine the evolution of the bridge width $w_0$ in the same terms as that of its height $h_0$. If the problem was perfectly decoupled into independently evolving height and width, one might expect inertial growth rates $h_0 \sim t^{2/3}$ and $w_0 \sim t^{1/2}$, \TS{the latter because in the $y-z$-plane projection the droplets exhibit initially a contact angle of $180^\circ$ (corresponding to $\theta=90^\circ$)}. 

In reality, however, a complex coupling between the two directions is to be expected, which is difficult to model with analytical approaches. The impact of their mutual influence on the bridge growth dynamics is an open question and one of the main topics addressed in this investigation.
 
Computer simulations are in principle a formidable tool to overcome experimental and analytical limitations, but accessing both regimes is not an easy task due to the wide range of surface tension and viscosity required~\cite{Datadien-Toschi2021,Espanol-Warren2017,Liu-Harting2016}. Furthermore, due to the intrinsic multiscale nature of coalescence, one has to resolve orders of magnitude in length scales to describe the system from the small initial bridge height to the full droplet size and beyond, including the surrounding hydrodynamic flow field~\cite{Janssen-Anderson2011,KampKraume2017}. 

In this article, we investigate the coalescence dynamics of liquid lenses using the color gradient lattice Boltzmann simulation method~\cite{Latva-Kokko-Rothman2005, Reis-Phillips2007, Leclaire-Reggio-Trepanier2013, Montessori-Succi2018}. This method overcomes some of the limitations of the pseudopotential lattice Boltzmann approach of Shan and Chen used in previous works~\cite{Shan1995,Liu-Harting2016,XH18a,HXH17,Datadien-Toschi2021}, which was not able to attain the viscous regime. The color gradient method allows us to cover both regimes by spanning more than four orders of magnitude in surface tension and more than two orders of magnitude in viscosity.

The remainder of the paper is organized as follows. In section II we introduce the color gradient lattice Boltzmann method, while section III and IV summarize our simulations of two coalescing top-down symmetric liquid lenses in 2d and 3d. The final section provides conclusions and a short outlook on future work.

\section{Lattice Boltzmann color gradient method}
Our simulations are conducted with the lattice Boltzmann method on a three-dimensional lattice with 19 discrete velocities (D3Q19)~\cite{benzi1992lattice}. The evolution of the discrete distribution function $f_i^k(\vec{x},t)$ for each fluid component $k$ is described by the lattice Boltzmann equation
\begin{equation}
f_i^k(\vec{x}+\vec{c}_i \Delta t, t + \Delta t) = f_i^k(\vec{x},t) + \Omega_i^k(\vec{x},t),
\end{equation}
where $\Omega_i^k$ is the collision operator, $i=1,...,19$ specifies the lattice direction and $k \in \{1,2,3\}$ the fluid component. In the following, we set the time step $\Delta t=1$ and the lattice constant $\Delta x=1$ for the sake of clarity without loss of generality. The fluid density $\rho^k$ is obtained from the zeroth moment of the distribution function
\begin{equation}
\rho^k(\vec{x},t) = \sum_i f_i^k(\vec{x},t),
\end{equation}
and (in absence of external forces) the macroscopic fluid velocity  $\vec{u}^k(\vec{x},t)$ from the first moment of the distribution function
\begin{equation}
\vec{u}^k(\vec{x},t) = \frac{ \sum_i f_i^k(\vec{x},t) \, \vec{c}_i}{\rho^k(\vec{x},t)}.
\end{equation}
To model phase separation we employ the color gradient method (CG) which introduces a coupling between the fluid components and performs the phase separation in three steps \TS{\cite{Gunstensen1991,Leclaire-Reggio-Trepanier2013}}: first, the color gradient, i.e. the direction of steepest increase in the density of the respective fluid component, is calculated
\begin{equation}
\vec{F}^k(\vec{x},t) = \nabla \left( \frac{\rho^\zeta(\vec{x},t) - \rho^\xi(\vec{x},t)}{\rho^\zeta(\vec{x},t) + \rho^\xi(\vec{x},t)} \right),
\end{equation}
where $\zeta,\xi \in \{1,2,3\} \;\; \textrm{and} \;\; \zeta>\xi$.

In the next step, also known as perturbation step, the populations that are collinear to the gradient of the color field are increased, while those perpendicular to it are decreased, resulting in the appearance of a surface tension term:
\begin{equation}
\left( \Omega_i^k \right)^\mathrm{pert} f_i^k (\vec{x},t) = f_i^k(\vec{x},t) + \frac{A_k}{2} |\vec{F}^k(\vec{x},t)| \left( w_i \cos^2(\phi_i^k) - B_i\right).
\label{CG_pert}
\end{equation} 
Here, $w_i$ are the lattice weights
\begin{equation}
w_i=\left\{%
\begin{array}{ll}
    1/3 & i=1 \\
    1/18 & i=2,...\,,7 \\
    1/36 &i=8,...\,,19 \\
\end{array}%
\right.
\end{equation}
and $\phi_i^k$ is the angle between the color gradient $\vec{F}^k$ and the lattice direction $\vec{c}_i$. $A_k$ is a free parameter determining the surface tension and $B_i$ is chosen as to ensure mass conservation:
\begin{equation}
 B_i=\left\{%
\begin{array}{ll}
    -2/9& i=1\\
    1/54& i=2,...\,,7\\
    1/27& i=8,...\,,19\\
\end{array}%
\right.
\end{equation}
 
Finally, the recoloring step separates two phases by distributing the two components to opposite directions
\begin{equation}
\left( \Omega_i^\zeta \right)^\mathrm{recol} f_i (\vec{x},t) = \frac{\rho^\zeta}{\rho} f_i(\vec{x},t)  + \beta \frac{\rho^\zeta \rho^\xi}{\rho^2} \cos(\phi_i) \sum_{k=\zeta, \xi} f_i^{k,eq}(\vec{x},t)(\rho^k,0),
\label{recoloring}
\end{equation}
where $\beta$ is a free parameter controlling the interface thickness ($\beta=0.99$ in all our simulations), $f_i=\sum_k f_i^k$ and $f_i^{k,eq}$ is the local equilibrium distribution derived from a Taylor expansion of the Maxwell-Boltzmann distribution to the second order
\begin{equation}
f_i^{k,eq}(\vec{x},t) = \rho^k \left[ \phi_i^k + \varphi_i \bar{\alpha} + w_i \left( \frac{\vec{c}_i \cdot \vec{u} }{c_s^2} + \frac{(\vec{c}_i \cdot \vec{u})^2 }{2 c_s^4} - \frac{\vec{u}^2 }{2 c_s^2} \right) \right],
\end{equation}
with $c_s$ being the lattice speed of sound\TS{, $\varphi_i$ a lattice dependent weight and $\bar{\alpha}$ the density weighted average of parameter $\alpha_k$ setting the equilibrium density for each fluid component \cite{Leclaire-Latt2017}}.
The total collision operator of the CG method $\Omega_i^k$ is an extension of the standard Bhatnagar-Gross-Krook (BGK) collision operator \cite{Bhatnagar-Krook1954}
\begin{equation}
\left( \Omega_i^k \right)^\mathrm{BGK} f_i^k(\vec{x},t) = f_i^k(\vec{x},t) - \omega_k \left( f_i^k(\vec{x},t) - f_i^{k,eq}(\vec{x},t) \right),
\end{equation}
which relaxes the population $f_i^k$ to its local equilibrium with a relaxation rate $\omega_k= 1/\tau_k$. From the Chapman-Enskog expansion to second order one can derive the relation between relaxation time $\tau_k$ and kinematic viscosity $\nu_k$ of fluid $k$ as 
\begin{equation}
\nu_k = c_s^2 \left( \tau_k- \frac{1}{2}  \right).
\end{equation}
Finally, the BGK operator is extended by the perturbation and recoloring operators to yield the CG collision operator $\Omega_i^k$,
\begin{equation}
\Omega_i^k= \left( \Omega_i^k \right)^\mathrm{recol}\circ\left( \Omega_i^k \right)^\mathrm{pert} \circ \left( \Omega_i^k \right)^\mathrm{BGK},
\end{equation}
which applies in a chain the BGK, perturbation and recoloring operators, in this order, and conserves all collisional invariants like mass and total momentum for each fluid component.
\TS{Differences in density of the fluids are taken into account by the parameter $\alpha_k$ determining the equilibrium density of each fluid component $k$ in the equilibrium distribution~\cite{Leclaire-Latt2017}. }\\

\TS{Our implementation has been carefully validated and demonstrated to properly reproduce, e.g., Neumann angles, the equation of Young-Laplace and the behavior of oscillating droplets -- in line with the data presented in~\cite{Leclaire-Reggio-Trepanier2011}.} 
\TS{The color gradient lattice Boltzmann method is a diffuse interface method and has as such some advantages in comparison to methods which require a distinct tracking of the fluid interfaces, as it is the case with the front-tracking~\cite{Tryggvason-Jan2001}, volume of fluid~\cite{Hirt-Nicholds1981} or level set~\cite{Osher-Sethian} methods. 
In our approach, interfaces arise naturally based on local interaction rules and thus complicated algorithms for the tracking of interfaces or computations of e.g. the local interface curvature as for instance in the volume of fluid method can be avoided~\cite{BRH16}. Phase separation, as well as nucleation and growth of droplets and bubbles appear naturally and do not require any adhoc treatment. Diffuse interface methods also allow for a straightforward implementation of wetting boundaries and an explicit treatment of contact lines can be avoided. Conceptionally, the CG LB method differs from other diffuse interface LB methods because the interaction rules are not clearly based on the underlying physical behaviour. For example, the pseudopotential method of Shan and Chen uses a phenomenological forcing term as a bottom up mean field description of molecular
interactions effectively leading to surface tension between two fluids~\cite{Shan1993}. The free energy LB method, on the contrary, is a top down model and based on the minimization of a free energy functional following the ideas of Cahn and Hilliard~\cite{SwiftYeomans1995}.
The main advantage CG method is that it is numerically more stable and allows for a much wider spectrum of surface tensions and viscosities. For example, too low surface tensions render fluids to become miscible in the pseudopotential and free energy methods, while interfaces are stable by definition for almost arbitrarily low values of $\sigma$ in the CG method. Even density and viscosity contrasts are easily treated within moderate limits~\cite{Leclaire-Reggio-Trepanier2013,Leclaire-Latt2017}. See a recent review of Liu et al.~\cite{Liu-Harting2016} and a comparison of the pseudopotential and CG method~\cite{Datadien-Toschi2021} for more details.
}

\section{Liquid lens coalescence in 2d}

Our study focuses on the coalescence of two identical, top-down symmetric liquid lenses. We begin our investigation with the quasi two-dimensional case (cylindrical symmetry), before later turning to the fully three-dimensional simulations. The droplets are initialized side by side and connected via a contact point (resolved by approximately 5 lattice nodes). Over time, surface tension drives the interface to minimize the surface area, and a bridge develops, which grows until the two droplets have merged into a single larger one.

The dynamics of the coalescence process is determined by the initial geometry of the droplets~\cite{Eddi-Snoeijer2013} and the combined effect of inertia, surface tension $\sigma$, and dynamic viscosity $\mu = \rho \nu$, where $\nu$ is the kinematic viscosity. These quantities determine a characteristic velocity scale, also known as capillary velocity, given by the ratio $v_c = \sigma / \mu$. The Reynolds number of the coalescing droplets can thus be expressed as $\textrm{Re}=\rho \, \sigma \, h_0 / \mu^2$~\cite{Eggers-Lister-Stone1999}. At early times the system is dominated by viscous forces, since the bridge height $h_0$ is much smaller than the viscous characteristic length $l_v = \mu^2/(\sigma \rho)$ \cite{Lee-Yarin2012}. In this regime $\textrm{Re} \ll 1$ and the flow is described by the Stokes equation. The crossover between the viscous and inertial regime occurs at $Re \approx 1$. From then on, viscous dissipation becomes increasingly negligible and the dynamics of the system is determined by inertial forces.

For small contact angles, the drop height is much smaller than its lateral extension which allows to apply the lubrication approximation, under which the Navier-Stokes equations simplify to yield the thin-sheet equation \cite{Erneux-Davis1993}
\begin{eqnarray}
    h_t + (u h)_y =& 0\\
    \rho (u_t + u u_y) =& \sigma \, h_{yyy} + 4 \mu \, \frac{(u_y h)_y}{h}.
\end{eqnarray}
By solving the thin-sheet equation with the similarity ansatz
\begin{equation}
    h(y,t)=k t^\alpha \mathcal{U}(\xi), \;\;\; u(y,t)= \frac{\alpha k}{\theta} t^\beta, \;\;\; \xi=\frac{\theta y}{k t^\alpha},
\end{equation}
it has been shown that the growth of the bridge between two coalescing lenses exhibits a power-law behavior with two asymptotic regimes~\cite{Hack-Snoeijer2020}. In the viscous regime, where viscous forces dominate inertial forces ($\rho \approx 0$), the bridge height grows linearly in time, $h_0(t) \sim t$, whereas in the inertial limit $h_0(t) \sim t^{2/3}$. The two asymptotic regimes as well as the crossover region can be described by the universal curve
\begin{equation}
h_0/h_c= \left( \frac{1}{t/t_c} + \frac{1}{(t/t_c)^{n}} \right)^{-1},
\label{eqn:powerlaw-crossover}
\end{equation}
where, in this case, $n=2/3$ and $t_c$ and $h_c$ are the crossover time and height that provide a universal scaling law~\cite{Hack-Snoeijer2020}.

The large viscosity and surface tension range required to reach the viscous as well as inertial regime is a major challenge for numerical approaches~\cite{Coreixas-Latt2019}. So far, the viscous regime was not amenable to the very popular pseudopotential lattice Boltzmann method of Shan and Chen due to its numerical instabilities at low values of the surface tensions~\cite{Liu-Harting2016}. The color gradient lattice Boltzmann method, on the contrary, is stable over a much wider range of surface tension values~\cite{Datadien-Toschi2021}.

For the quasi two-dimensional case we perform simulations of a domain consisting of $4 \times 2048 \times 768$ lattice points in $x$,$y$ and $z$ direction (pseudo 2d) with periodic boundary conditions \TS{(a minimal domain length of 4 lattice nodes is required by our specific implementation)}. The droplets are initialized with a radius of $282$ lattice nodes and
a contact angle $\theta=30^{\circ}$. The distance of the droplet edges to the periodic domain boundaries are chosen sufficiently large such that their mutual influence across the periodic boundaries can be neglected. Each lens was previously equilibrated separately in its surrounding fluid, making sure that the lenses are initially at rest and have no initial velocity of approach. 

\TS{We perform a series of simulations by varying the droplet surface tension ($\sigma=[0.000058;0.116]$) and the surface tension between the two surrounding fluids ($\sigma_o=[0.0001;0.2]$) over three orders of magnitude as well as the droplet viscosity over one order of magnitude ($\mu=[0.01;0.5]$) to yield low and high capillary velocities, respectively. The viscosity of the outer fluids ($\mu_o=[0.00166;0.02]$) is chosen to be as small as possible while still in the stable regime of the LB algorithm ($\tau>0.5$). This results in an overall change in the $Re$ number of 5 orders of magnitude, allowing us to investigate both the viscous and inertial regime.}

Furthermore, to collapse the bridge growth for different capillary velocities on a single master curve, we use $t_c=\frac{288 K_i}{K_v^3}\frac{\mu^3}{\rho \sigma^2 \theta^2}$ and $h_c= \frac{72 K_i}{K_v^2}\frac{\mu^2}{\rho \sigma}$ with $K_i=0.106$ and $K_v=2.21$ as previously obtained from similarity solutions of the thin-sheet equation~\cite{Hack-Snoeijer2020}. Since we are only interested in the initial phase of the coalescence to limit finite size effects, we stop our simulations when $h_0$ has reached $2/3$ of the height of the lenses.  

\begin{figure}
    \begin{center}
        \includegraphics[width=0.61\columnwidth]{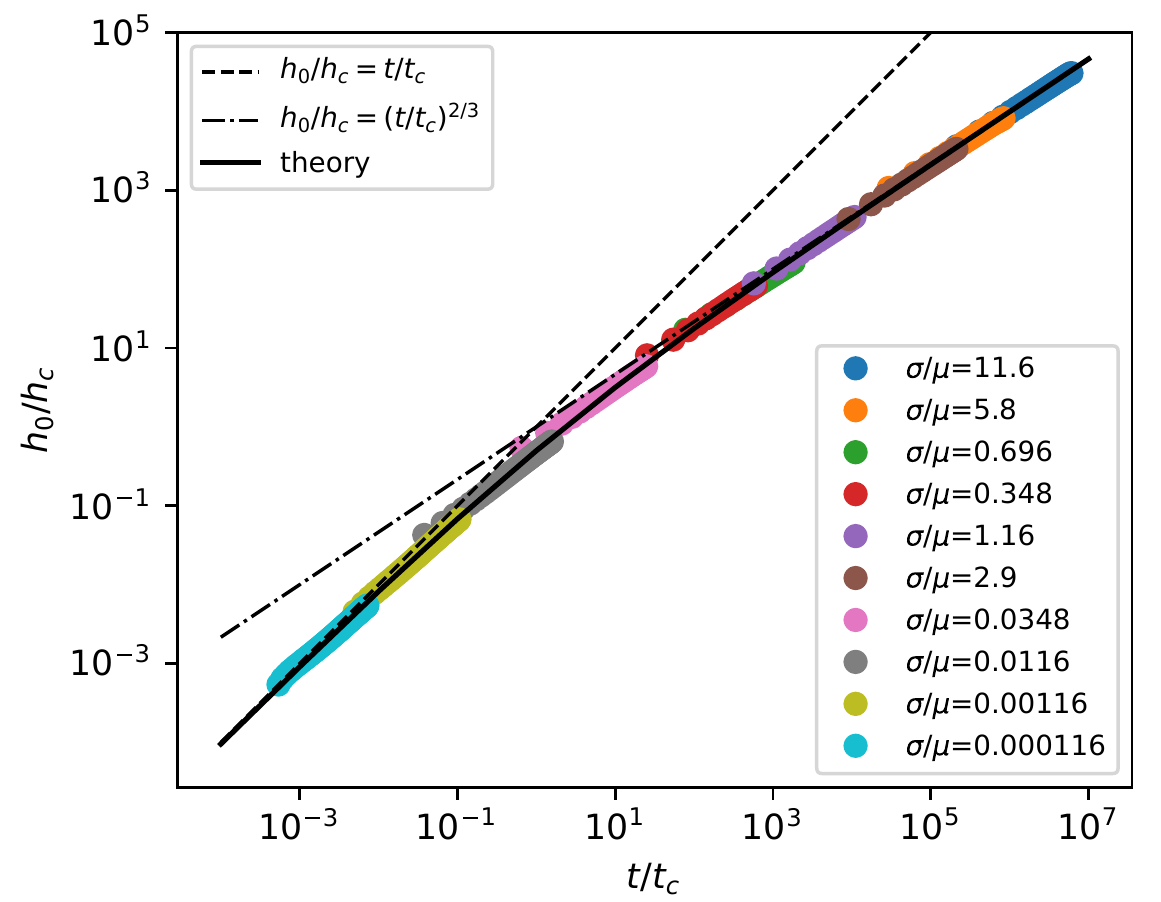}\hspace*{1.1cm}
    \end{center}
\caption{Power law relation for the bridge growth in $2d$ covering the viscous as well as inertial regime (solid line: interpolation according to Eq.~(\ref{eqn:powerlaw-crossover}), dashed line: viscous theory, dotted-dashed line: inertial theory). \TS{Note that due to the double logarithmic representation, the initial bridge height $h_0($t=0$)=0$, which is the same for each data set, is not shown.}
}
\label{fig:scaling}    
\end{figure}

In the viscous regime our simulations yield a linear bridge growth $h_0 \sim t$, followed by a crossover region that provides a smooth transition towards the $h_0 \sim t^{2/3}$ dependence of the inertial regime (see Fig.~\ref{fig:scaling}). All simulations show very good agreement with the analytical solution of the thin-sheet equations. Noticeably, the numerical constants $K_i$ and $K_v$ from~\cite{Hack-Snoeijer2020} yield an excellent collapse of the data sets, confirming that the thin-sheet equation is a good approximation to describe the coalescence dynamics in the case of small contact angles.   

\begin{figure}
     \centering
         \includegraphics[width=.49\columnwidth]{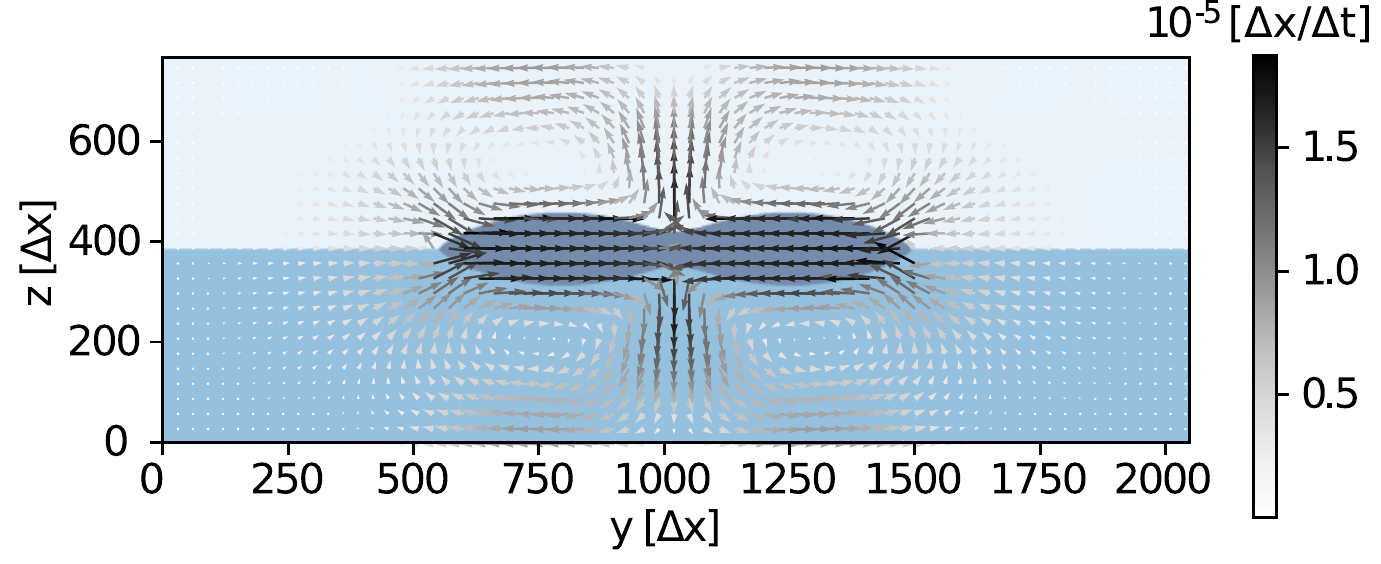}
         \includegraphics[width=.495\columnwidth]{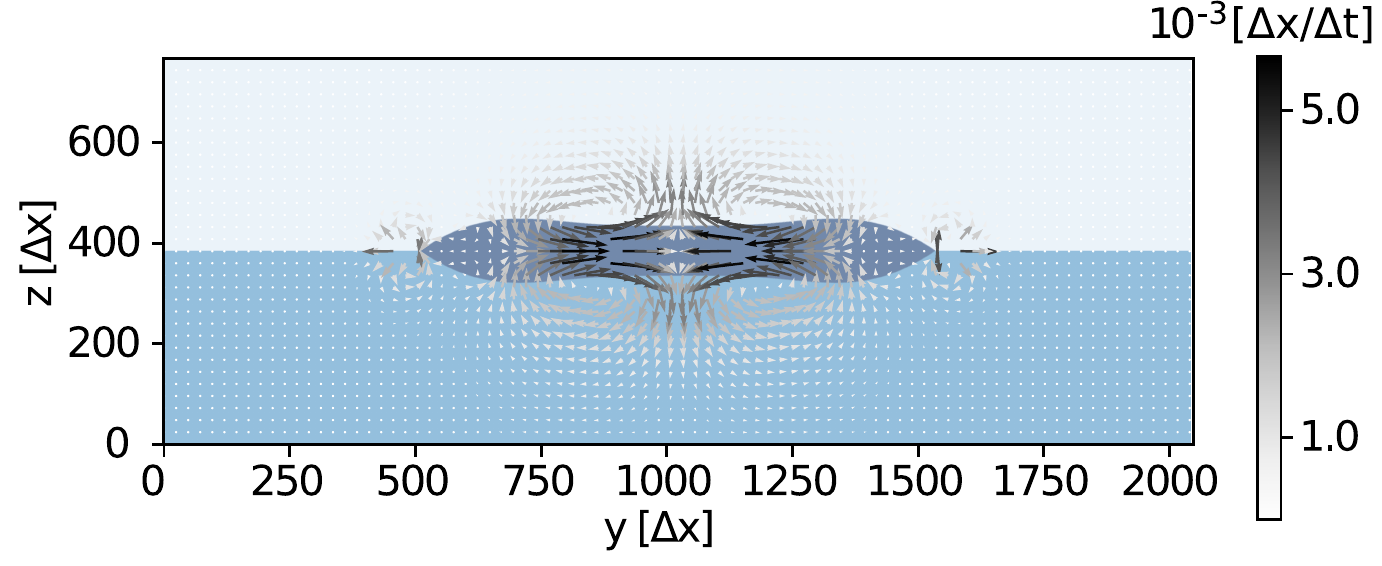}
     \caption{Flow field of viscous ($\sigma/\mu=0.000116$, left) and inertial ($\sigma/\mu=0.348$, right) liquid lens coalescence, where the grey scale of the velocity vectors represents the magnitude of the velocity vectors. 
     }
\label{fig:VelField}
\end{figure}

The velocity field in the viscous regime is inherently dipolar and approaches a plug flow inside the liquid lens phase over time -- see Fig.~\ref{fig:VelField} (left panel) for a representative velocity field obtained from the simulations. While in the vicinity of the bridge minimum the flow field of the inertial regime is still dipolar (Fig.~\ref{fig:VelField}, right panel), two additional dipolar flow structures arise approximately at the center of each of the two initial liquid lenses. Furthermore, at larger distances from the bridge center fluid inertia causes the appearance of circulations in the wake of the retracting tips of the liquid lenses. 

\begin{figure}
     \centering
         \includegraphics[width=.495\columnwidth]{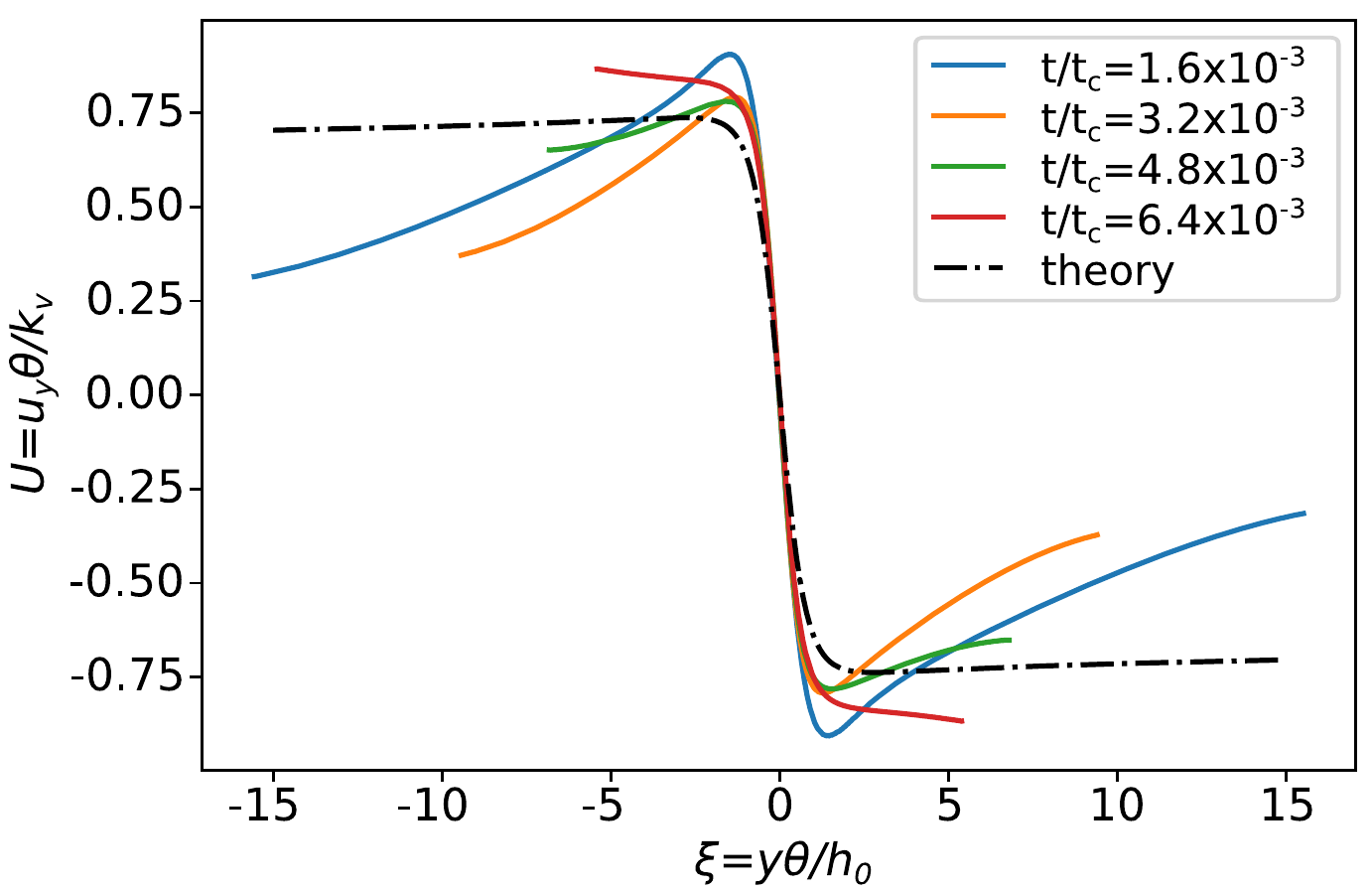}
         \includegraphics[width=.495\columnwidth]{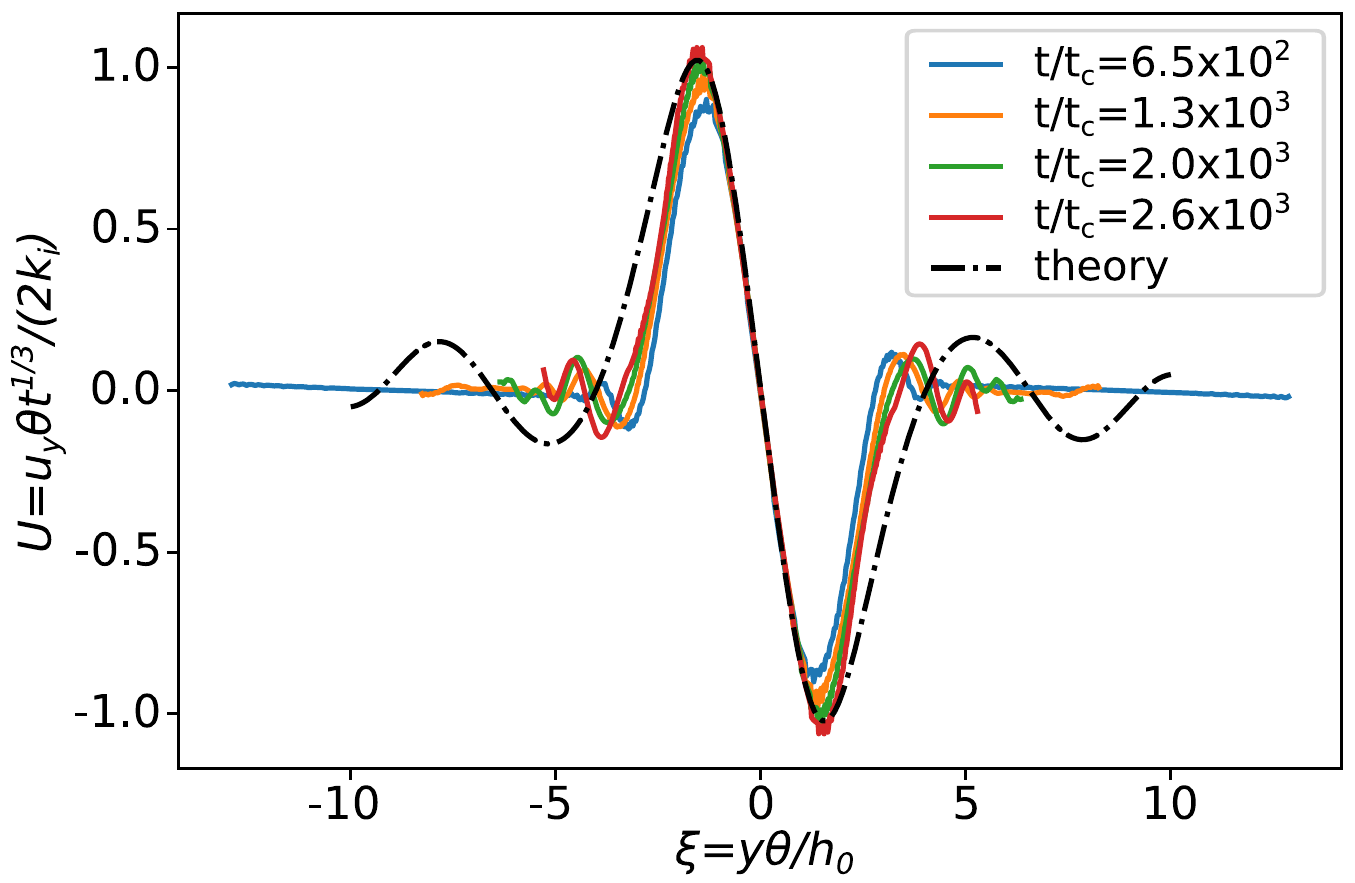}
     \caption{Average profile of the $y$-component of the velocity at different times in the viscous (left) and  inertial (right) regime compared to thin-sheet theory. 
     }
     \label{fig:VelProf_temporal_full_y}
\end{figure}

In analogy to the assumptions of the thin-sheet equation, Fig.~\ref{fig:VelProf_temporal_full_y} shows the profile $u_y(y,t)$ of the $y$-component of the velocity, averaged over the droplet extension along the $z$ axis. Close to the bridge center ($|\xi|<1$) the velocity profile is in good agreement with the prediction of the thin-sheet equation for the viscous as well as the inertial case. At larger distances to $h_0$ {($|\xi| > 1$)}, however, the simulated velocity profile starts deviating from the thin-sheet solution. This effect can be attributed to the finite size of the lens as well as the difference in the treatment of the outer fluids: In contrast to the thin-sheet equation, our simulations include the full dynamics of the surrounding fluids with a finite viscosity. Thus, viscous damping in the surrounding fluids influences the velocity field inside the droplets.  

\begin{figure}[ht]
     \centering
         \includegraphics[width=0.49\columnwidth]{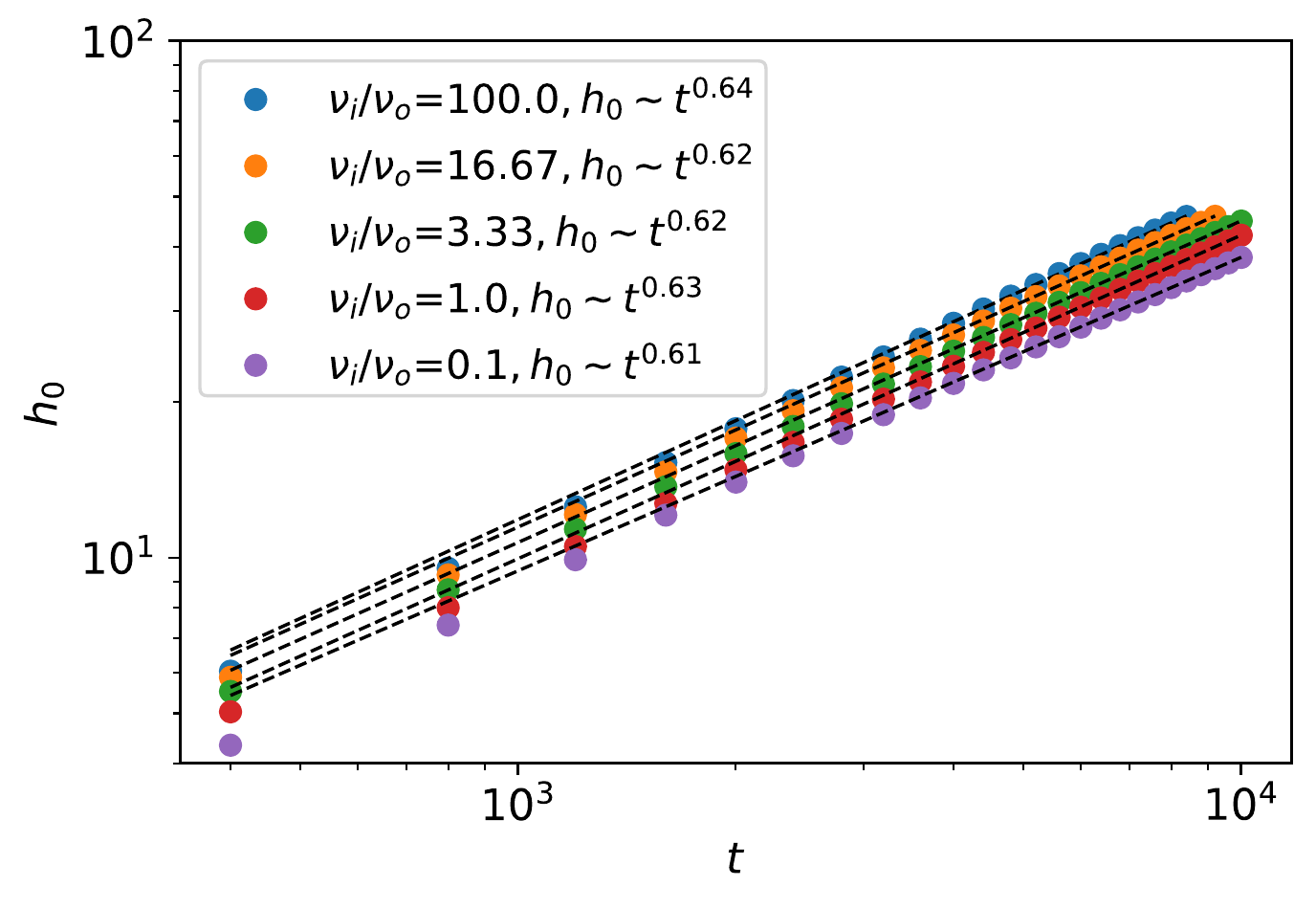}
         \includegraphics[width=0.49\columnwidth]{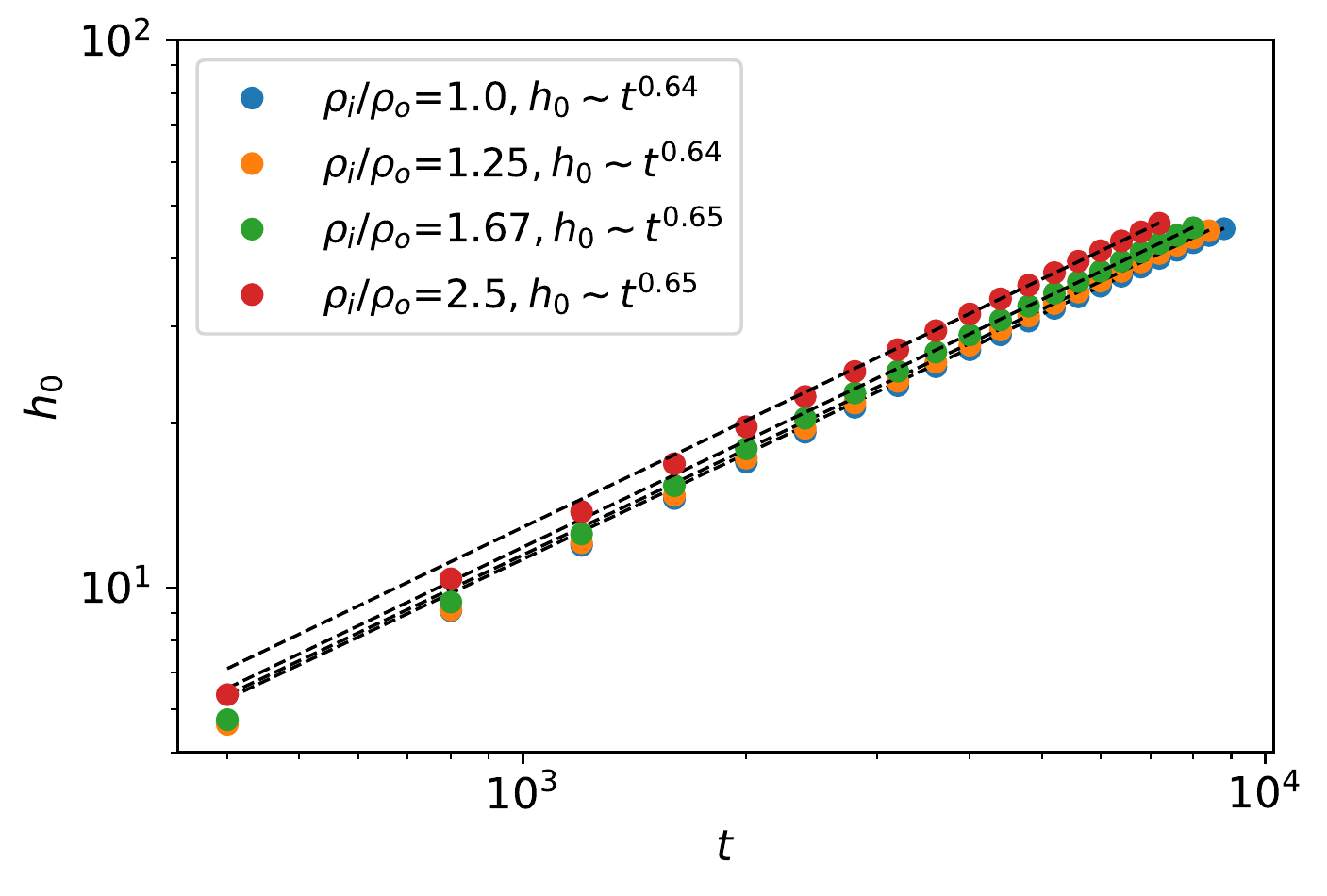}
     \caption{\TS{Influence of the variation of outer fluid viscosity $\nu_o$, where the viscosity of the liquid lens $\nu_i=1/6$ and density ratio $\rho_i/\rho_o=1$ are fixed (left). Influence of the variation of density ratio $\rho_i/\rho_o$, where inner density $\rho_i=1$ and viscosity ratio $\nu_i/\nu_o=100$ are fixed (right)}. 
     }
\label{fig:ViscAndDensVar}
\end{figure}

\TS{To be able to compare our simulation results to the similarity solution of thin-sheet theory we ensured that the coalescence process is dominated by the flow inside the liquid lenses by choosing the fluid viscosity of the outside fluid to be at least one order of magnitude smaller than that of the lenses. We verified that the external fluid does not alter the scaling characteristics of the bridge growth within the range of parameters studied (see Fig.~\ref{fig:ViscAndDensVar}). Although we varied the viscosity ratio of the inner to the outer fluid $\nu_i/\nu_o$ by three orders of magnitude, the scaling exponent changes only to a small extent ($\approx 4\%$). Similarly, for a $2.5$-fold increase of the outer fluid density, the scaling exponent varies only by $\approx 1\%$.}

\section{Liquid lens coalescence in 3d}

\begin{figure}
    \centering
  \includegraphics[width=.8\columnwidth]{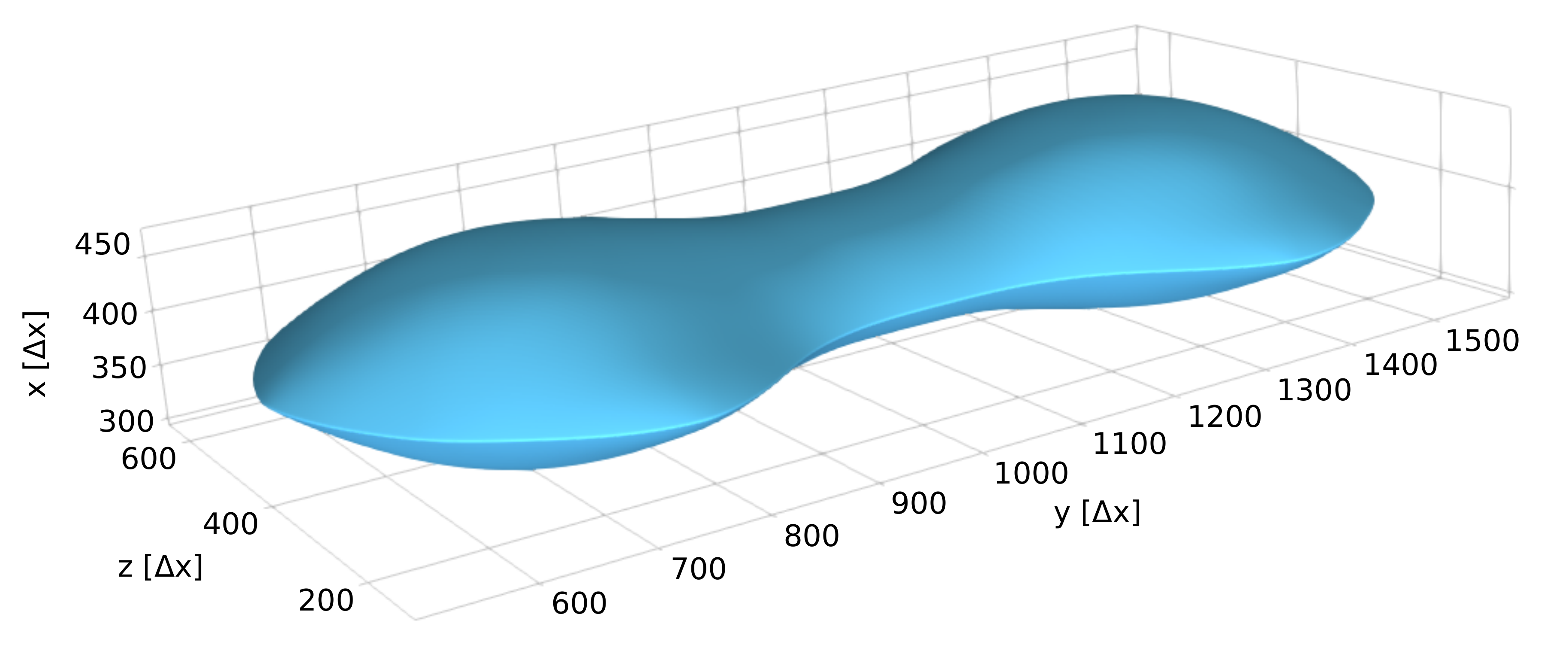}
    \caption{Snapshot of two coalescing liquid lenses in 3d. The snapshot is taken at $t/t_c=285.8$ (12,000 $\Delta t$), where the connecting bridge has already developed {for a capillary velocity $\sigma/\mu=2.9$ (inertial regime)}.
    }
    \label{fig:3d_2drops}
\end{figure}

Next, we extend our simulations to the fully three-dimensional case (Fig.~\ref{fig:3d_2drops}), where we use a system size of $768 \times 2096 \times 768$ lattice nodes in $x$, $y$ and $z$ direction with periodic boundary conditions. The update of $1.2 \cdot 10^{9}$ lattice sites requires a considerable amount of computational resources. Therefore, the simulations were conducted on the JURECA Booster machine with $32,768$ Intel KNL cores using up to $3.4$ million core-hours to generate a single data set. 

\begin{figure}
  \centering
  \includegraphics[width=.495\linewidth]{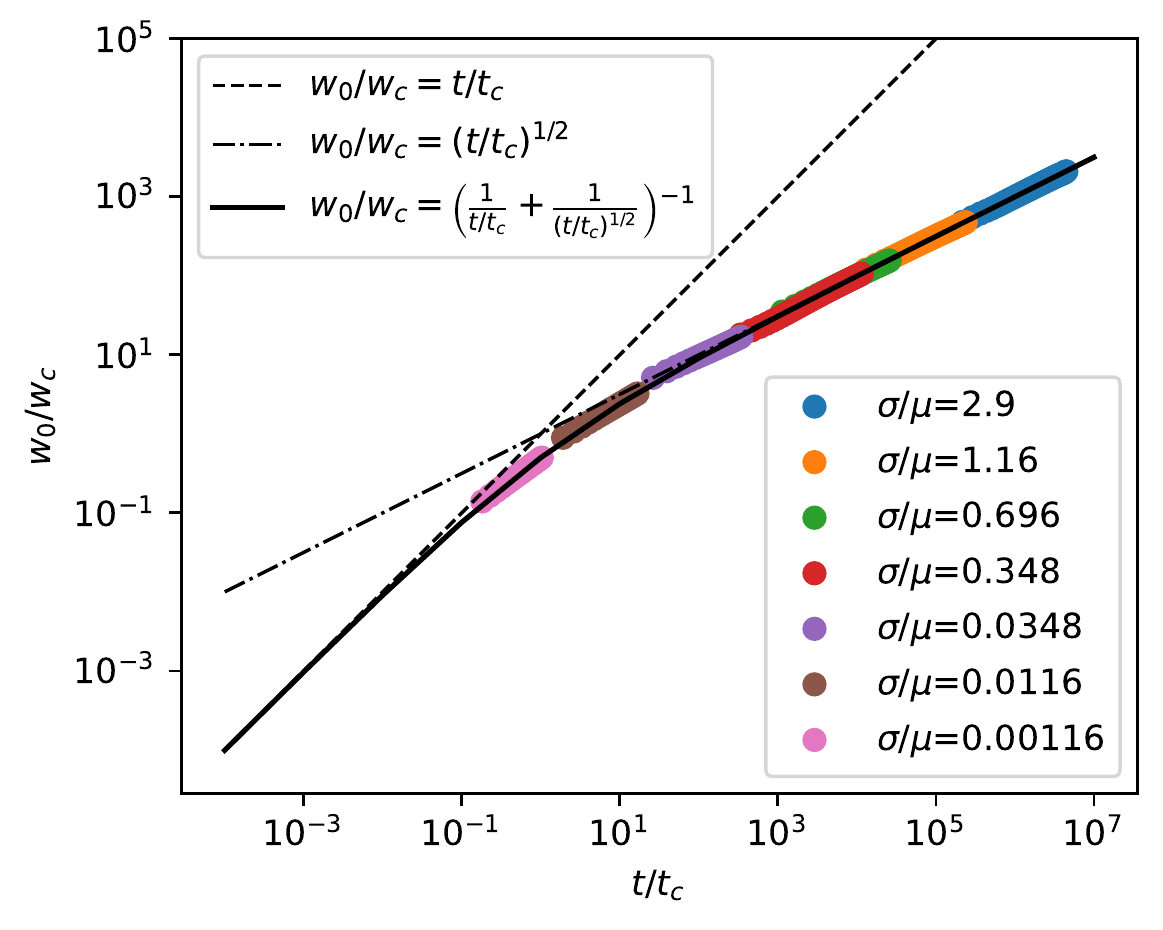} 
  \includegraphics[width=.495\linewidth]{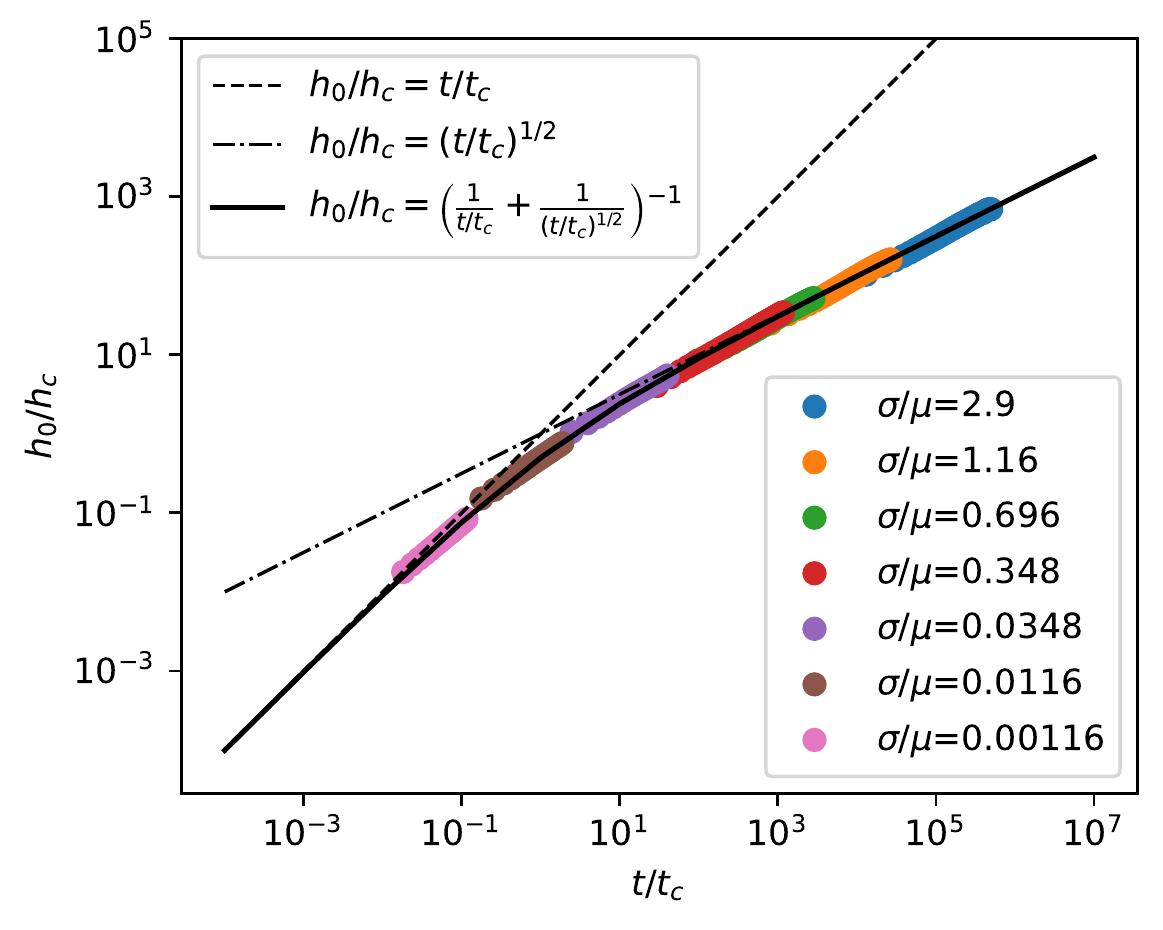} 
\caption{Power law relation for the bridge growth in $3d$ covering the viscous as well as inertial limit (solid line: interpolation according to Eq.~(\ref{eqn:powerlaw-crossover}), dashed line: $t$, dotted-dashed line: $t^{1/2}$). Left panel: bridge width $w_0(t)$; right panel: bridge height $h_0(t)$. \TS{Note that due to the double logarithmic representation, the initial bridge height $h_0($t=0$)=0$, which is the same for each data set, is not shown.}}
\label{fig:scaling3d}
\end{figure}

In analogy to the pseudo two-dimensional case, we initialize two equilibrated lenses with a contact angle of $\theta=30^{\circ}$ (see Fig.~\ref{fig:profiles3d}) and adequate spacing to the domain boundaries.
The growth of the bridge width reported in the left panel of Fig.~\ref{fig:scaling3d} scales as $w_0 \sim t^{1/2}$, which agrees with experiments~\cite{Aarts-Bonn2005, Burton-Taborek2007, Paulsen-Nagel2011, Eddi-Snoeijer2013}, analytical~\cite{Eggers-Lister-Stone1999, Xia-Zhang2019} and numerical studies~\cite{Wang-Sun2018, Montessori-Succi2019} for freely suspended, respectively spherical droplets.
The evolution of the bridge height $h_0$, on the contrary, does not behave as in the quasi two-dimensional case ($t^{2/3}$ scaling), but follows again the scaling $h_0 \sim t^{1/2}$ found for the width, as reported in the right panel of Fig.~\ref{fig:scaling3d}. This indicates that the thin-sheet equation for the 2d case fails to describe the dynamics of the three-dimensional bridge growth. The scaling law is however not in contradiction to the experimental data shown in Ref.~\cite{Hack-Snoeijer2020}, 
where reasonably the transition region between the viscous and the inertial regime was observed.
In the three-dimensional case the naive assumption of a decoupled width and height growth is clearly not satisfied. Since the two directions are strongly coupled, it is reasonable to expect that $w_0$, which entails a larger amount of fluid than $h_0$, is dominating the dynamics of the inertial regime for the whole bridge.

\begin{figure}[t]
\begin{center}
    \includegraphics[width=0.6 \columnwidth]{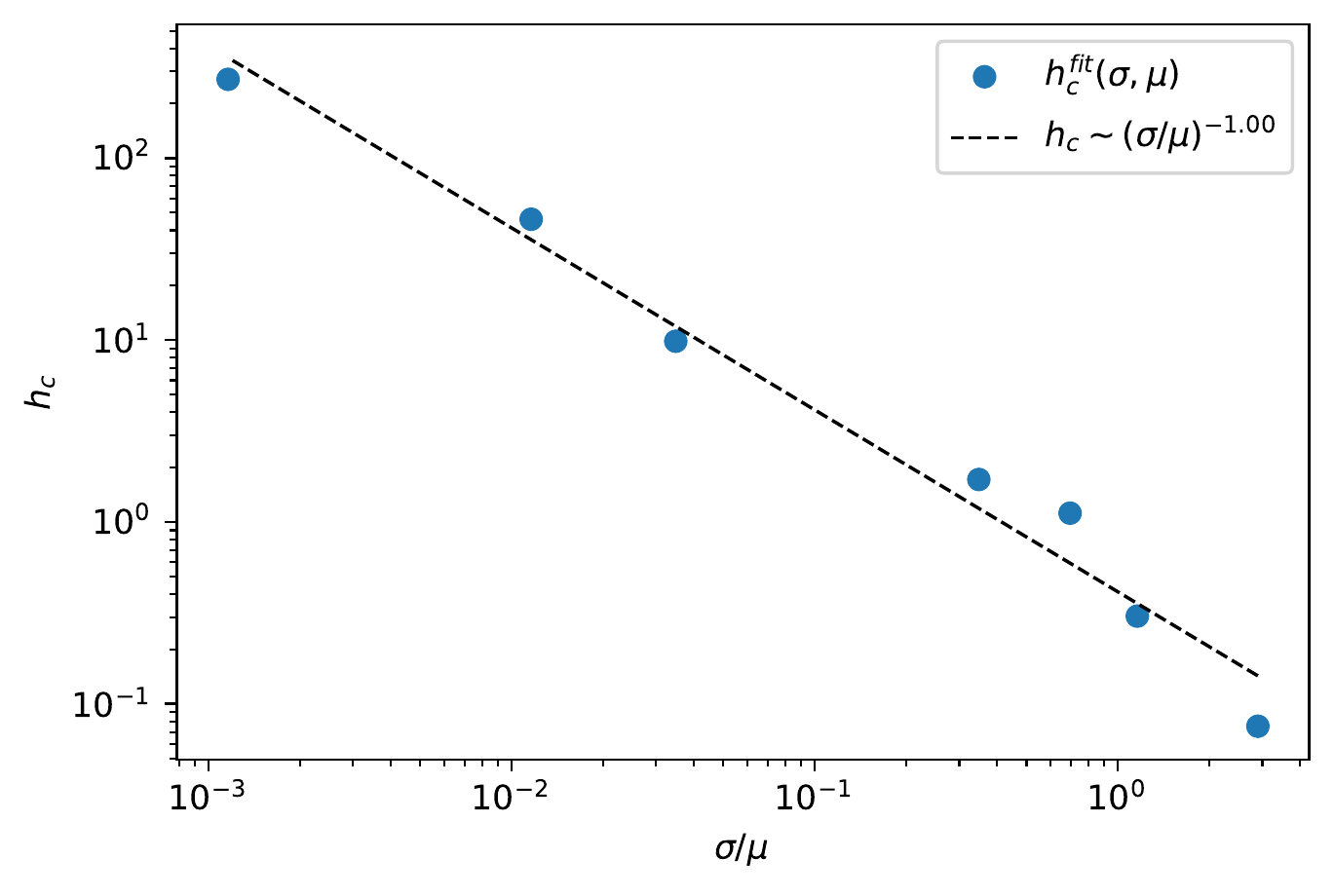} 
    \caption{Dependence of the best-fit $h_c$ on the capillary velocity in $3$d. The dashed line represents the linear relation obtained by fitting the exponent of capillary velocity ($\sigma/\mu$) to the data points.
    }
    \label{fig:hc_fit3d}
\end{center}
\end{figure}

In this case, we could not use $h_c$ as predicted by the analytical solution of the thin-sheet equation, and we settled for finding the best fitting value of $h_c$ for each data set. To check that the solution is not arbitrary, we plot the values of $h_c$ as a function of the ratio $\sigma/\mu$ of each data set, as reported in Fig.~\ref{fig:hc_fit3d}. The dependence is clearly of the type $h_c\sim \mu/\sigma$. However, since $h_c$ can be expressed dimensionally in terms of surface tension and viscosity as $h_c\sim \mu^2 / (\sigma \rho)$, it is clear that this relation incorporates a (constant) prefactor with the dimensions of a kinematic viscosity. 

\section{Conclusion}

Liquid lens coalescence is an intrinsically multiscale problem and studying its scaling laws involves investigating surface tensions and viscosities that cover several orders of magnitude. Our simulation method - the color-gradient lattice Boltzmann method - has proven to deliver hydrodynamically consistent results for the required wide parameter ranges. This allows us to investigate the coalescence dynamics from the viscous to the inertial regime. For the pseudo two-dimensional case we find good agreement with the similarity solutions of the thin-sheet equation. In the viscous regime the bridge grows linearly with time and in the inertial regime, the bridge growth is proportional to $t^{2/3}$. 

\TS{The three-dimensional coalescence simulations, on the contrary, deviate from the similarity solution of the thin-sheet equation in the inertial regime. Here, a $t^{1/2}$ dependence replaces the $t^{2/3}$ scaling.} 
This can be explained by a strong coupling between the two directions and the involvement of a larger mass of fluid in the bridge width as compared to the bridge height. This renders the dynamics of the bridge width the dominant process.

These results underline the necessity of a more generic theoretical framework for a more accurate understanding of the general coalescence process. In future studies, the influence of asymmetric properties of the liquid lenses on the coalescence dynamics could be investigated, for instance by extending the simulations to top-down asymmetric lenses or lenses with different viscosities or even non-Newtonian properties.

\begin{acknowledgments}
We acknowledge Jacco Snoeijer and Michiel Hack for fruitful discussions.
This work has received financial support from the Deutsche Forschungsgemeinschaft (DFG, German Research Foundation),  within the priority program SPP2171 ``Dynamic Wetting of Flexible, Adaptive, and Switchable Substrates'', projects HA-4382/11-1 and SE-3019/1-1 as well as SFB 1452 ``Catalysis at liquid interfaces'', Project-ID 431791331. We also thank the J\"ulich Supercomputing Centre for providing the necessary computing time.
\end{acknowledgments}


%

\end{document}